\documentclass{aa} 

\usepackage{psfig}

\newcommand{\der}[3]{\frac{d^{#3} {#1}}{d {#2}^{#3}}}

\renewcommand{\Re}{\rm Re}
\renewcommand{\Im}{\rm Im}

\newcommand{\tbuo}{t_{\rm buo}}

\newcommand{\comega}{C_{\Omega}}
\newcommand{\tg}{t_{\rm G}}
\newcommand{\td}{t_{\rm D}}

\newcommand{\Bphi}{B_\varphi}
\newcommand{\Bphiz}{B_{\varphi0}}
\newcommand{\Br}{B_r}
\newcommand{\Brz}{B_{r0}}
\newcommand{\Myr}{{\rm \ Myr}}

\authorrunning{Lesch \& Hanasz}
\titlerunning{Cosmic ray dynamo in very young galaxies}

\begin{document}

\title{Strong magnetic fields and cosmic rays in very young galaxies}

\author{H. Lesch \inst{1}
\and
M. Hanasz \inst{2}}

\institute{
University Observatory, M\"unchen University, Scheinerstr. 1, D-81679, Germany and 
Center for Interdisciplinary Plasma Science (CIPS),
\email{lesch@usm.uni-muenchen.de}
\and 
Toru\'n Centre for Astronomy, Nicholas Copernicus University
PL-87148 Piwnice/Toru\'n, Poland, \email{mhanasz@astri.uni.torun.pl}
}

\date{Received 24 September 2002/ accepted 7 January 2003}
\setcounter{footnote}{0}

\abstract{
We present a scenario for efficient magnetization of very young galaxies about
0.5 Gigayears after the Big-Bang by a cosmic ray-driven dynamo. These objects
experience a phase of strong star formation during this first $10^9$ years. We
transfer the knowledge of the connection between star formation and the
production rate of cosmic rays by supernova remnants to such high redshift
objects. Since the supernova rate is a direct measure for the production rate
of cosmic rays we conclude that very young galaxies must be strong sources of
cosmic rays. The key argument of our model is the finding that magnetic fields
and cosmic rays are dynamically coupled, i.e. a strong cosmic ray source
contains strong magnetic fields since the relativistic particles drive an
efficient dynamo in a galaxy via their buoyancy.   We construct a
phenomenological model of a dynamo driven by buoyancy of cosmic rays and show
that if azimuthal shearing is strong enough the dynamo amplification timescale
is close to the buoyancy timescale of the order of several $10^7 \div 10^8$
yr. We predict that young galaxies are strongly magnetized and may contribute
significantly to the gamma-ray-background.
\keywords{galaxies: evolution - ISM: magnetic fields - cosmic rays}
}
\maketitle

\section{Introduction}

Faraday rotation measurements of high redshift objects ($z>2$) clearly indicate
that these galaxies have micro Gauss fields correlated over spatial scales of
several kpc ($> 5 \div 10$) at least (Athreya et al. 1998, Carilli \& Taylor 2002
and references therein). Regardless of the specifics of the Faraday screen
model, the intrinsic Rotation Measures (RM) of several thousand rad${\rm
m^{-2}}$ require such strong fields although the universe was just a sixth of
its present age at redshifts $z=2.25$ and only a tenth at $z=3.8$. It is
surprising that fields at such early times are similar to those found in
galaxies and clusters today. These observations pose a considerable challenge
to the models of magnetic field generation because of the small time available
for amplifications at those redshifts, of the order of a few hundred million
years. This problem has been discussed in many publications (e.g. Lesch \&
Chiba 1995; Kronberg 1994 and references therein).

Here we discuss the magnetic field generation in terms of galaxy evolution at
high redshifts, taking into account the consequences of the first phases of
intense star formation in young galaxies. Despite many problems of the details
there is now a general consensus that about half a Gigayear after the Big-Bang
galaxies experience a phase of strong star formation (Steidel et al. 1999).
Whereas the massive galaxies with a stellar mass at
redshift $z=0$ larger than $M_{\rm Stellar}\geq 2\cdot 10^9\, h^{-1} M_\odot$
continue to form stars until the present epoch (Majewski 1993; Alonso-Herrero
et al. 1996) (h is the Hubble constant in units of $H_0 = 100\, {\rm km s^{-1}
Mpc^{-1}})$), less massive galaxies stop forming stars at higher redshifts
($z=2 \div 3$) (Grebel 1997, Matteo 1998).

Blain et al. (1999) reconstruct the star formation histories compatible with
the observations at various wavelengths incorporating SCUBA results (Hughes et
al. 1998) and derivations from chemical evolution at high $z$. Their favorite
model peaks at about $z=2$ and stays constant with higher redshift. Such a star
formation history was predicted by CDM simulations (Nagamine et al. 2001).
Likewise the comparison of the redshift evolution of the global star formation
rate (SFR) from such simulations with the observations is in good agreement.
The CDM model can explain the decreasing SFR with time. In other words, the SFR
is on average a smooth function and levels off at redshifts $z> 4$. This
behaviour is consistent with an exponential decay of the SFR in time, on a time
scale of about half a Gigayear.

Very recently Lanzetta et al. (2002) addressed the effects of cosmological 
surface brightness dimming on observations of faint galaxies. They come to the 
conclusion that the incidence of the highest intensity star formation regions 
increases monotonically with redshift. At the very end of their paper they stated 
"Our analysis suggests that star 
formation in the very early universe may have occurred at a much higher rate than 
is generally believed and that cosmological dimming effects cannot be ignored when
interpreting statistical properties of the high-redshift galaxy population"

In any case in the early phases of galaxy formation, several hundred million
years after the Big-Bang the Universe was populated with intensively star
forming galaxies. This is the background for our scenario about the rapid
magnetization of very young galaxies via cosmic ray driven-dynamo action.

By rapid and strong magnetization we mean the amplification of galactic
magnetic fields up to the equipartition values of the order of a few $\mu {\rm
G}$ on a timescale of $10^8$ yr starting from the beginning of star formation
period in early galaxies. 

One can estimate that the energy output from supernovae in the form cosmic rays  is
more than sufficient. Assuming one SN per year we get $10^8$ SN exploding within
$10^8$ yr. Assuming that 10 \% of SN kinetic energy output is converted into the
cosmic ray energy we obtain $10^{58}$ erg of energy released in the form of cosmic
rays within $10^{8}$ yr.

For a typical disk volume of the order of $300 \,{\rm kpc}^{3}$, a uniform magnetic
field  of the strength  5 $\mu {\rm G}$ contains $10^{55}$ ergs of magnetic energy.
This is 3 orders of magnitudes less than the cosmic ray energy output within $10^8$
yr. We note however that during the bursty phase of galactic evolution losses of
both cosmic rays and magnetic field from galactic disks should be significant. 

It is the aim of our contribution to show that the scenario of intense star
formation at high redshifts has profound implications for the magnetic energy
content of a galaxy. We will show that the connection between star formation
and magnetic field generation and amplification is provided by the cosmic rays.

\section{Cosmic Rays and the FIR-Radio-Correlation}

Our first subject concerns the existence of cosmic rays in galaxies. Where
do they come from? What can we learn from our Galaxy and galaxies in
the local universe? How can we extrapolate our findings into the
early phases of galaxy evolution?

Since cosmic rays (CR) in galaxies consist mainly of protons and nuclei which
do not emit much radiation, the main information about CR comes from their
relativistic electron component which  emit radio synchrotron radiation in the
galactic magnetic fields. The relativistic electrons are accelerated in
supernova remnants (Koyama et al. 1995, Reynolds 1996, Parizot \& Drury 2000). 
The observational evidence for this
acceleration scenario is overwhelming since disk and starburst galaxies exhibit
a very tight correlation between the spatially integrated nonthermal radio
continuum flux at 6cm and the thermal far-infrared (FIR)- flux at wavelengths
between 40 and 120 $\mu {\rm m}$ (Dickey \& Salpeter 1984; Helou et al. 1985;
Wunderlich, Klein \& Wielebinski 1987; Wunderlich \& Klein 1988; Condon \&
Broderick 1988; V\"olk 1989). This well-known radio-FIR-correlation has a
dispersion of only 0.2 in the logarithm of the ratio of both fluxes over more
than four orders of magnitudes (Wunderlich, Klein \& Wielebinski 1987).

Since the FIR emission is mainly starlight from OB-stars absorbed and
reradiated by dust, the qualitative interpretation of the correlation was that
the nonthermal emission in the radio band must be related to recent star
formation and in fact that relativistic electrons responsible for the
synchrotron emission should originate mainly in supernova remnants. Thus, the
radio-FIR-correlation connects the star formation and magnetized cosmic rays by
particle acceleration in supernova remnants. This is the first piece of
evidence that star formation is accompanied by intense cosmic ray production.

A second piece of evidence was presented by V\"olk, Klein \& Wielebinski
(1989). They argued on the basis of the spallation age of galactic cosmic
rays and the radio luminosity that the supernova remnants are the main
sources for the cosmic rays. They applied this argument to the starburst
galaxy M82 and our Galaxy and found the very surprising result that for both
galaxies the simple relation holds that the cosmic ray production rate $\dot
E_{\rm CR}$ is proportional to the supernova rate $f_{\rm SN}$: $\dot E_{\rm
CR} \propto f_{\rm SN}$. This is a surprise since both galaxies are
completely different with respect to their cosmic ray energy density,
magnetic energy density and supernova rate $f_{\rm SN}$ (in our Galaxy
$f_{\rm SN} =(30\, {\rm yr})^{-1}$, whereas in M82 $f_{\rm SN}\simeq (1\,
{\rm yr})^{-1}$ (Schaaf et al. 1989)). Especially cosmic ray and magnetic
energy density differ by two orders of magnitudes, in M82 they are 100 times
larger than in our Galaxy. V\"olk, Klein \& Wielebinski (1989) concluded that
their simple scaling should be applicable for all kinds of galaxies. In other
words, in all galaxies, no matter what evolutionary state they have reached
the supernova rate is a direct measure for the cosmic ray production rate.

The total cosmic ray injection power per unit volume is 
\begin{equation}
{dW_{CR}\over{dt}}= \left<{{Eq}}\right>V={ 
\left<{{Eq}}\right>\over{\bar\rho}}\bar M,
\end{equation}
where $\bar M=\bar \rho V$ is the gas mass contained in the cosmic ray 
propagation volume $V$. $\left<{{Eq}}\right>$ denotes the cosmic ray injection power per unit volume
\begin{equation}
\left<{{Eq}}\right>=A\int^{\infty}_0  E\, q_E\, dE,
\end{equation}
with the number $q_E$ of accelerated particles per unit volume, unit time and 
unit energy band. The mass number $A$ accounts for the fact that $E$ is the energy 
per nucleon.  Although the cosmic ray power depends on the source 
spectrum via $q_E$, on the gas mass and the propagation volume, different 
authors find quite an agreement in absolute numbers. Drury et al. (1989) as well 
as Fields et al. (2001) found a total cosmic ray power for the Galaxy
\begin{equation}
{dW_{CR}\over{dt}}= 1.1- 1.5\times 10^{41}\, {\rm erg\, s^{-1}},
\end{equation}
using an injection spectrum proportional to $p^{-2.2}$ taken from the 
observations. 

As mentioned above there is no doubt that shock waves originating  from
supernova explosions play a major role for the acceleration of cosmic rays 
(for electron acceleration to energies above 100 TeV see e.g. Koyama et al.
1995;  Tanimori et al. 1998, for direct evidence of nucleon acceleration to  
$E > 300 {\rm MeV/nucleon}$ see Combi et al. 1998). The net Galactic cosmic ray 
injection power due to supernovae ${dW_{SN}\over{dt}}$ is proportional to the 
supernova rate and  average mechanical energy output of a supernova times some
efficiency which is transformed into cosmic rays. Since the supernova rate is
proportional to the star formation rate  ${dW_{SF}\over{dt}}$ one finally ends
up with a scaling that  ${dW_{SN}\over{dt}}\propto {dW_{SF}\over{dt}}$. Fields
et  al. (2001) derived this intuitive scaling from quantitative arguments by a 
careful and consistent model for the energetics of cosmic ray acceleration and 
the abundances of LiBeB, which are due to cosmic ray nucleosynthesis in the
interstellar medium. Doing so, they could go a step further and they derive  an
even more direct scaling of the star formation rate and the cosmic ray 
acceleration rate, as has been implicitly assumed by most previous work on 
early Galactic cosmic ray nucleosynthesis (see next section). 

Here we can summarize that beyond any 
doubt the rate at which stars are formed in a galaxy is a direct measure for 
the cosmic ray flux.

\section{Arguments from the scenario of Galactic cosmic ray nucleosynthesis}

An interesting side-line of our scenario concerns the observations of Be and B
in old stars of the Milky Way halo (Gilmore et al. 1992; Duncan, Lambert \&
Lemke 1992). B and Be are spallation products of cosmic ray protons colliding
with C, N and O. The observations prove that these elements were present in the
early Galaxy in amounts much larger than conventional scenarios of
nucleosynthesis  can account for. The B/Be-ratio is, however, quite consistent
with the Galactic cosmic rays (GCR) nucleosynthesis scenario. In order to
explain this intriguing results Prantzos, Casse \& Vangioni-Flam (1993) and
Prantzos \& Casse (1995) showed that one solution is that during the early
phase of the Galaxy GCR were more efficiently confined than today. As a result
very high GCR fluxes were obtained. If this is a generic feature of the early
phases of galaxy evolution, large gamma-ray fluxes resulting from pion decay
should have been produced at redshifts $z >3\div10$. Prantzos \& Casse  also
found that the GCR in the early phases of Galaxy evolution must have been more
confined to the Galaxy than today to establish the B/Be ratio.  This means that
something kept the GCR closer to the Galaxy as today. This can only be provided
by a stronger magnetic field and/or a higher gas density. Latter is reasonably
to expect in young galaxies. Both components are acting on the CR. It is
currently thought that GCR are accelerated mainly by supernovae shock waves and
propagate in the Galaxy mainly by diffusing on the irregularities of the
Galactic magnetic field, thereby suffering energy losses by ionization, nuclear
reactions and leakage in the framework of the "leaky box" model (e.g., Cesarsky
1980).  A higher confinement  at earlier phases than means that the box was
"closed" rather than a "leaky" box for GCRs (Prantzos \& Casse 1995). The
component of the interstellar medium which can hinder the GCR to escape freely
is the magnetic field. 

This requirement must be understood in terms of a significantly higher
supernova rate than today, since Parizot and Drury (2000) showed that the
elements B and Be can be produced in the early Galaxy only,  by the combined
action of collective supernova exploding in OB-associations driving expanding
superbubbles.  That is, the high amount of B and Be in the old stars tells us
that many more supernova explosions must have been occurred at an early
evolutionary stage of the Galaxy to provide these elements. Or simply, that the
supernova rate at the early stages was much higher than today. If we combine
these results with those from V\"olk, Klein \& Wielebinski (1989) the physical
origin of an enhanced CR-flux in the early phases of the Galaxy has to be a
higher supernova rate and a higher magnetic field strength than today. Somehow
there must be a dynamical coupling between the supernova remnant shock waves
the CR and the magnetic field strength.

\section{Cosmic Rays and Dynamo Action}

Thus,  we now introduce the connection of the magnetic field to the
radio-FIR-correlation, which was found by Lisenfeld, V\"olk \& Xu (1996). They
presented an analysis of the radio-FIR-correlation in starburst galaxies in
comparison to a control sample of normal spiral galaxies. Here we cite their
main conclusions:

- The mean FIR to radio ratio showed no significant difference between the
starburst sample and the control sample and is independent of the starburst
strength.

- The magnetic field is required to be amplified strongly on short time-scales
($\sim 10^7\, {\rm yr}$) in order to maintain a sufficiently constant FIR to
radio flux ratio in the first $2\times 10^7  {\rm yr}$ of a starburst.
Otherwise the inverse Compton losses during this phase would lead to a very low
synchrotron emission, resulting in a too large value of the FIR to radio flux
ratio, significantly higher than the observed one.

What can we learn from these findings for the high redshift objects? We can
learn that an intense starburst must produce lots of CR and must produce strong
magnetic fields to confine the CR to explain the B/Be-ratio. We know the sources
for the cosmic rays - the supernova remnants. What is the source for the field
amplification? Our answer is: the CR population.

In the Milky Way the radial distribution of the molecular clouds where stars
are born, the distribution of supernova remnants and the distribution of cosmic
rays are different. The column density of molecular gas comprises a peak at the
Galactic Center  and a  peak at the galactocentric radius $R_{\rm G}=4.5 {\rm
kpc}$. Behind $R_G= 10 {\rm kpc}$ the column density drops out below a few per
cent of its peak value (Clemens et al 1988). The supernova distribution has a
peak at 4.5 kpc and extends up to 14 kpc (Case and Bhattacharya 1998). The
galactic cosmic rays have a considerably broader distribution up to 20-30 kpc
in galactic radius (Strong and Moskalenko 1998). The latter distribution
suggests  the existence of a self-regulation (valve) mechanism  which removes
the excess of cosmic rays above certain critical level of cosmic ray energy
density (see Hanasz and Lesch 2001, Ptuskin 2001). This supports the model of
Breitschwerdt et al. (1993) who showed the possibility of an intense cosmic
ray-driven galactic wind if the CR-energy density exceeds a critical value.

The supernovae influence the dynamics of ISM in multiple ways through the input
of kinetic energy, thermal energy and cosmic ray energy to the surrounding
medium. The kinetic energy injected in a certain spatial scale cascades to both
smaller and larger scales in the form of turbulent cascade. The thermal energy
resulting from the shock heating produces the X-ray emission of ISM and leads
to other observable effects like supernova remnants, superbubbles, galactic
chimneys (Korpi et al. 1999) and galactic fountains (Kahn and Brett 1993)

Cosmic rays are dynamically  important  for the ISM because their energy
density is comparable  to the energy densities of the remaining ingredients of
ISM. This means that  on average cosmic rays contribute to the pressure which
is comparable to the  pressure of thermal gas (including the thermal and
turbulent pressure) and to  the pressure of magnetic field (see e.g. Ferriere
1998a for a detailed model of the Galactic ISM). An additional factor essential
for the distribution of various ISM ingredients in galactic disks is gravity,
which comes partially from the mass of stars, dark matter and the thermal gas
itself. One should note that cosmic rays and magnetic field are weightless
components. Cosmic ray particles provide a significant pressure to the
multi-component interstellar medium even though their total mass in negligible
(1 cosmic ray nucleon per $10^9 {\rm cm}^3$). Since they contribute to the total
pressure but {\em do not} contribute to the mass density of ISM, they support
partially the heavy interstellar gas from below. Such a configuration is known
to be unstable against the buoyancy instability which is known as Parker
instability. The Parker instability operates in magnetized, gravitationally stratified
atmospheres  (Parker 1966, 1967) and is a magnetic version of the
Rayleigh-Taylor instability.

All the three chains of supernova energy conversion are potential sources of
magnetic field amplification, however the efficiency of each chain is
different.

 The kinetic energy is supposed to contribute to the cascade of helical
turbulence which plays a central role in the classical mean field dynamo
theory. About 10 percent of SN energy is converted to turbulent energy (Dorfi
1993). The hot gas contributes to the dynamo $\alpha$-effect through the
expansion of supernova remnants and superbubbles (Ferriere 1998b), or through
galactic fountains (Kahn and Brett 1993). Although majority of the  SN energy
becomes thermal energy at some stage of SNR evolution, there are two chains of
losses of the thermal gas. The first one is cooling of the hot interstellar
gas, the second is the escape of the light and buoyant hot gas from galactic
disks in chimney flows.

However, the hot gas fills only less than 20 \% of the present day Milky Way
ISM volume (Ferriere, 2001), mostly in hot bubbles. The hot bubbles are
supposed to be the source for the $\alpha$-effect in conventional dynamos 
(Ferriere and Schmidt 2000), which are by a factor of 10 too slow for the
amplification of the regular magnetic field component (Lesch \& Camenzind
1994, Hanasz \& Lesch 2001)

On the other hand, cosmic rays are much more efficient since they fill the
whole galaxies (due to the diffusive propagation) and couple dynamically to the
magnetic field. The dynamical coupling of the cosmic ray component and magnetic
field follows from the fact that cosmic ray particles can propagate only along
magnetic field lines (the gyroradius of typical 1 GeV protons in the $\mu{\rm
G}$ magnetic field is of the order of $10^{13}$ cm, which is 8 orders of
magnitude below the typical galactic spatial scales. Moreover, the bulk
propagation of cosmic rays along magnetic field lines is slowed down much below
the speed of light. Due to the streaming instability cosmic rays excite
turbulence (Kulsrud and Pearce 1969). The turbulence scatter cosmic ray
particles back and forth and the bulk motion of relativistic cosmic ray gas is
slowed down to the Alfv\'en speed, which is typically a few tens km/s.

Parker (1992) proposed a new kind of galactic dynamo powered by the very
powerful output of cosmic rays accelerated in supernova remnants.  That kind of
fast dynamo incorporating the buoyancy of cosmic rays, Coriolis force,
differential rotation  and magnetic reconnection is expected  to amplify the
large scale magnetic field in timescales  $\leq 10^8 {\rm yr}$ or even
faster.   We are going to demonstrate in the next section  that the
amplification timescale of the cosmic ray driven dynamo is indeed related to
the buoyancy timescale, which is of the order of $\leq 10^8$ yr.

It was demonstrated by Hanasz \& Lesch (2000) that the Parker instability
excited by the cosmic ray supply in supernova remnants becomes nonlinear
immediately after explosion, resulting in a significant acceleration of the
Parker instability above the standards of the linear approximation. The same
circumstance - nonlinearity since the very beginning implies that all the
linear instability criteria become invalid. The Parker (buoyancy) instability
is usually considered as a destructive effect for large scale galactic magnetic
fields (see Hanasz \& Lesch 2001),  however in the presence of Coriolis  force,
differential rotation and magnetic reconnection cosmic rays provide a means for
a very efficient amplification of the large scale magnetic field. 

The idea of Parker's (1992) {\em fast galactic dynamo} is based on the sequence
of  following processes. The primary driver of this kind of dynamo in galactic
disks is the powerful supply of cosmic rays in supernova remnants. The cosmic
rays inflate lobes on the magnetic field. The magnetic lobes extending up to
the galactic halo are tightly packed and form tangential discontinuities in
magnetic field. Then fast magnetic reconnection starts to change the topology
of magnetic field forming either closed magnetic loops as it is proposed in the
original Parker's scenario, or with the aid of Coriolis force, helical tubes of
magnetic field. (We note following Parker that the term {\em fast dynamo} is
due to the contribution of fast magnetic reconnection, which provides the
necessary dissipation of the small scale magnetic field). The latter
possibility has been demonstrated by means of numerical simulation by Hanasz,
Otmianowska-Mazur and Lesch (2002). The further reconnection of loops (which
undergo rotation caused by Coriolis force) or helical tubes  leads to the
production of the radial component of magnetic field. The new radial magnetic
field is subsequently stretched by differential rotation, leading to the
amplification of the toroidal component of magnetic field. 

\section{Phenomenological model of the buoyancy driven dynamo}

In the present section we construct a phenomenological buoyancy-driven dynamo
model resulting from our previous investigations  (Hanasz \& Lesch 1998, 2000,
2002 and Hanasz, Otmianowska-Mazur \& Lesch 2002).  The buoyancy (Parker)
instability operates in the absence of cosmic rays, however if cosmic rays are
additionally present in the system, the buoyancy timescale shrinks
significantly. Investigations of buoyancy driven galactic dynamos were
performed also by other authors  (Moss et al. 1999) incorporate two
modifications to the standard dynamo equation. First, they assume that the
dynamo $\alpha$ parameter is an increasing function of magnetic field strength
and second, that the large-scale rotation velocity is supplemented by a regular
outflow velocity due to buoyancy, however they did not take into
account cosmic rays. In their investigations efficient amplification of the
large-scale magnetic field is possible if magnetic field responsible for
buoyanacy is strong enough. In the current model buoyancy is provided by both
magnetic field and cosmic rays, so the strength of magnetic field is 
not crucial. Our model relies on the following three elements:

\begin{enumerate}

\item In the absence of rotation, buoyancy of the magnetic field and cosmic
rays  evacuates the magnetic field from the disk on a timescale of a few 100
Myr,  without the cosmic ray contribution, down to a few 10 of Myr when the
Parker instability is driven by cosmic ray injection in supernova remnants. We 
define the buoyancy timescale $t_{\rm buo}$ as the characteristic time of removal
of the magnetic field via Poynting flux (instead of relating it to the growth time
of Parker instability in the linear regime). In our model the vertical Poynting
flux replaces the concept of turbulent diffusivity in classical turbulent
dynamos.

\item It follows from a linear analysis of Parker instability and from
numerical simulations (Hanasz \& Lesch 1998) that the instability leads to the
formation of kpc-scale helices on initially horizontal magnetic field lines.
This finding is a modification of Parker's (1992) original scenario involving
the dynamics of closed magnetic loops. The formation of helices, which result
from the action of the Coriolis force and  tidal forces in differentially rotating
disks, is equivalent to the production of a negative radial magnetic field
component in lower layers of the disk and a positive magnetic field component
in upper layers of the disk. This scenario has been confirmed by numerical
simulations presented in recent papers  (Hanasz\& Lesch 2000, 2002; Hanasz,
Otmianowska-Mazur \& Lesch 2002). Hanasz \& Lesch (1998) have shown that the
production of the radial magnetic field component in the helices is equivalent to
the presence of  the dynamo $\alpha$-effect. The neighboring helices merge as a
result of fast magnetic reconnection, producing a large-scale poloidal magnetic
field. {\em The essential point of our argument is that, according to 
numerical simulations, in the presence of realistic (rigid) rotation, the radial
magnetic field  of the significant magnitude of  $ 10 \div 30 \%$ (larger and
smaller values are possible) of the azimuthal component is  built within the
buoyancy timescale $t_{\rm buo}$}.

\item We take into account differential rotation which produces the azimuthal
magnetic field from the poloidal one. The differential rotation enters our
system through the shearing parameter $G=r d\Omega/dr$. The timescale of
generation of an azimuthal magnetic field from a poloidal one is $\tg =
1/|G|$

\end{enumerate}

Assuming a thin disk geometry, the superposition of all three mentioned effects
can be expressed by the following set of equations describing evolution of the
mean magnetic field $B_r$ and $B_\varphi$ in the disk

\begin{equation}
\der{\Br}{t}{} = - \frac{\comega}{\tbuo}{} \Bphi - \frac{1}{\tbuo}{} \Br,
\label{br}
\end{equation}

\begin{equation}
\der{\Bphi}{t}{} = -\frac{1}{\tg} \Br+ \frac{\comega}{\tbuo}{} \Br
- \frac{1}{\tbuo}\Bphi, \label{bphi}
\end{equation}
where $\comega\sim 0.1\div 0.3$ is a constant factor representing our
expectation that radial magnetic field of the order of $0.1\div 0.3$ of the
azimuthal magnetic field strength is efficiently produced within the timescale
$\tbuo$.   The constant factors $\pm\comega$ are related to the dynamo $\alpha$
parameter as described by  Hanasz \& Lesch (1998). 

The present dynamo concept relies on the assumption that fast magnetic
reconnection efficiently merges neighboring helices and produces a new
large-scale poloidal magnetic field of opposite sign in the lower half and the
upper half of the disk thickness $H$. The mean vertical buoyant velocity
removes the magnetic field from the upper layer of the disk volume. After
$\tbuo$ only the magnetic field from the lower layer  remains within the disk
volume. 

The negative sign in front of $\comega$ in Eqn. (4) and positive sign in
front of $\comega$ in Eqn. (5) are justified by the fact that the lower
(remaining) parts of spirals are twisted by the Coriolis force in the direction of
galactic rotation while the upper (rejected) parts are twisted against the
galactic rotation (see Figs.~1. and  2. in Hanasz \& Lesch 1998).

We search for harmonic solutions of the system of equations
(4) and (5) through the substitution
\begin{equation}
\left(\matrix{\Br \cr \Bphi}\right) =
\left(\matrix{\Brz \cr \Bphiz}\right) \exp (-i \omega t) 
\end{equation}
and obtain the following equation for the complex oscillation frequency: 

\begin{equation}
\omega^2 + \frac{2 i}{\tbuo} \omega + \frac{\comega}{\tg \tbuo}
-\frac{\comega^2}{\tbuo^2} - \frac{1}{\tbuo^2}=0 .
\label{disprel} 
\end{equation}
Complex solutions of the above second order polynomial equation are

\begin{equation}
\omega_\pm = \frac{1}{\tbuo}\left[-i \pm 
\left(\comega^2-\comega\frac{\tbuo}{\tg}\right)^{1/2}\right] .
\end{equation}
The real part of $\omega$ is the oscillation frequency and the imaginary part
represents the growth rate of the large-scale magnetic field. We note that in
the absence of rotation, only a damped non-oscillating solution is present, ie.
the growth rate is $\Im (\omega) = -1/\tbuo $. If rigid rotation ($\tg=0$) is
present then the real part of $\omega$ becomes non-vanishing and the damped
solutions  oscillate with frequency $\Re (\omega) = \pm \comega/\tbuo$.  If in
addition differential rotation is present and is strong enough ($\tg \leq
\tbuo/\comega$) then solutions become non-oscillatory and the rotation
contributes to the imaginary part of $\omega$. 

Amplification of the mean magnetic field is possible when $\Im (\omega_{+}) >0$.
The condition for the existence of growing solutions can be expressed as

\begin{equation}
\tg < \frac{\tbuo}{\comega + \frac{1}{\comega}} .\label{ampcond}
\end{equation}
The factor $1/(\comega+1/\comega)$ reaches its maximum value of 0.5 at 
$\comega=1$ so we deduce that the weakest restriction for
the shearing timescale coming around $\comega=1$ is
\begin{equation}
\tg < 0.5 \tbuo .
\end{equation}
We note that realistic present day galactic disks seem to fulfill this
relation, e.g. in the Milky Way $\tg\simeq 15-30 \Myr$ for galactocentric radii
ranging from  $0.5 R_\odot$ to $1 R_\odot$, $\tbuo$ is perhaps $50 \div
100\Myr$. 

If $\comega < 1$ then we find that amplification is possible if $\tg <
\tbuo/2.5$ for $\comega=0.5$, $\tg < \tbuo/3.5$ for $\comega=0.3$ and $\tg <
\tbuo/10$ for $\comega=0.1$. The limitation (10) for the shearing
timescale seems to be quite restrictive, however, different origins of shearing
motions can be considered: the galactic differential rotation, the shearing
resulting from the presence of galactic bars, density waves as well as galactic
close encounters.

The most essential property of our model is that {\em if only the amplification
condition (\ref{ampcond}) is fulfilled, the amplification of the mean magnetic
field proceeds on a timescale which is close to the buoyancy timescale}. The
buoyancy timescale $\tbuo$ is, on the other hand, controlled by the production
rate of cosmic rays. The presented dynamo model directly relates the dynamo
amplification timescale $\td$ 
\begin{equation}
\td = \frac{1}{\omega_{+}} = \tbuo \left[ 
\left(\comega\frac{\tbuo}{\tg}-\comega^2\right)^{1/2}-1\right]^{-1}
\end{equation}
to the production rate of cosmic rays. Although it is difficult to precisely
quantify this relation, we can say that the buoyancy timescale is at least an
order of magnitude shorter than the corresponding turbulent diffusion timescale
controlling the classical dynamos; thus, we deduce that the cosmic ray-driven
dynamo is an order of magnitude faster than the classical turbulent dynamos.

We expect therefore that the dynamo amplification timescale $\td$ may be very
short in galaxies intensely producing cosmic rays; however the amplification
condition requires that the shearing timescale $\tg$ is a few times shorter
than $\tbuo$. Finally, we note that the present phenomenological model is far
from precise, although it seems to display correctly the basic properties and
role of different timescales in our dynamo model.

\section{Implications}

The whole process  forming a closed cycle is expected to amplify the
large-scale galactic magnetic field on a time scale $\td\sim10^8$ yr as deduced
for the present-day galaxies like our Milky Way.   The fast dynamo process 
involves magnetic reconnection serving as a dissipative process  converting 
the smaller-scale turbulent field into the large scale field, so the turbulent
field has to form first within a few $10^7$ yr. The latter timescale
corresponds to the growth time of magnetic buoyancy instability. The detailed
field structure seems to be not very crucial to explain the findings by 
Lisenfeld, V\"olk \& Xu (1996); however the field structure is decisive for the
cosmic rays transport  (see eg. Giacalone and Jokipii 1999) and the onset of
Parker instability (Parker \& Jokipii 2000, Hanasz \& Lesch 2000). Strong
organized magnetic fields are observed in the radio halo of M82 (Reuter et al.
1992) and in compact dwarf galaxies (Klein, Weiland and Brinks 1991). Except in
the nuclear region of M82 the magnetic field is horizontal in significant parts
of the disk of that galaxy (Reuter et al 1992).  This confirms  the link
between the turbulent and the large-scale magnetic fields in starburst
galaxies.

Due to the significantly higher star formation rate in
early galaxies we expect that the strength of the primary driver - the cosmic
ray production-rate was much larger in the past leading to even faster
amplification of galactic magnetic fields. We present the idea of a fast
cosmic ray driven dynamo in the left part of diagram in Fig.~1.

\begin{figure}

\centerline{\psfig{figure=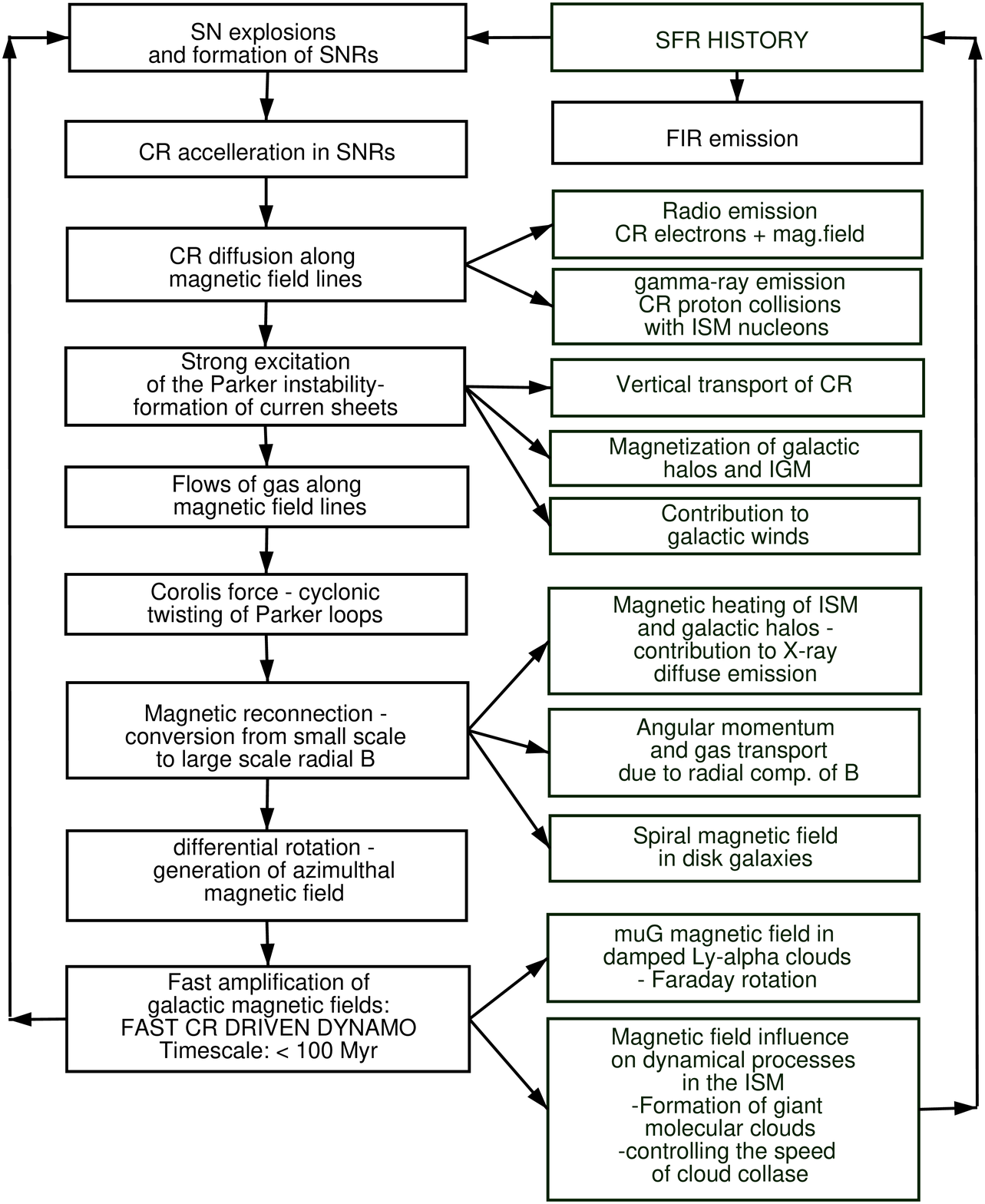,width=0.5\textwidth}}
\caption[]{Flow diagram of the cosmic ray driven fast dynamo and related
phenomena}
\end{figure}

In the right part of the diagram we point out various effects that are
physically related to the fast dynamo cycle. First, the star formation
history is the input parameter for the cosmic ray driven dynamo. The stellar
evolution leads to the production of dust in the interstellar medium, which is
the source of the FIR radiation. Second, the direct implication of the presence
of cosmic rays is the radio synchrotron emission from cosmic ray electrons
gyrating in the galactic magnetic field and the gamma-ray emission coming from
collisions of CR protons against ISM nucleons.

Third, the direct consequence of the cosmic ray driven Parker instability is
vertical transport of ISM components, primarily cosmic rays which cumulate
at tops of rising Parker loops (see Fig.~3 of Hanasz and Lesch 2000). The
vertical buoyant motions influence also the magnetic field which is very
efficiently removed from the disk by the Parker instability (see Fig.~8  of
Hanasz, Otmianowska-Mazur and Lesch 2002). The buoyant rise of magnetic
lobes and cosmic rays push gas just above, leading to so-called "helmet
flows", resembling the familiar bursts of solar wind.

Fourth,  magnetic reconnection converts the small-scale helical structures
coherently twisted by the Coriolis force into the large-scale spiral magnetic
field. The horizontal alignment of the resulting field forms the condition for
the next outburst of the Parker instability.  Moreover, magnetic heating of the
ISM and galactic halo has been proposed by Parker (1992) and further
investigated by Hanasz and Lesch (1998) and other authors (Tanuma et al. 2003).
The presence of a radial magnetic field has to give rise to some angular
momentum transport in galactic disks on cosmological timescales. The
reconnection and topological evolution of the distorted magnetic field forms
specific conditions for cosmic ray propagation. As we pointed out already, the
trajectories of cosmic ray particles are determined by the shape of magnetic
field lines, therefore after the onset of the Parker instability cosmic rays
are released from the disk (the vertical growth of magnetic loops with cosmic
rays does not stop) and after the reconnection/relaxation of the magnetic field
structure, cosmic rays are confined in the disk by the magnetic field.

Fifth, the fast magnetic field amplification offered by the fast dynamo model
provides a chance to explain the $\mu{\rm G}$ magnetic field in early galaxies.
(Wolfe, Lanzetta and Oren 1992).

Often it is stated that in radio galaxies at high redshifts the active radio
source itself  magnetizes the ambient thermal plasma. This was discussed by
Athreya et al. (1998) in  their conclusions: The synchrotron plasmas in the
radio lobes are permeated by strong magnetic fields of hundreds of $\mu$G or
even a magnitude higher in  the active hotspot, especially in powerful radio
sources. The magnetic field is  supposed to be effectively transported by the
radio jet from the radio nucleus.  Turbulent diffusion and
magnetohydrodynamical instabilities may lead to such   magnetic field transport
into the surrounding thermal plasma (Bicknell et al.  1990). However, there is
no observational evidence for such transport. The  observations show almost no
mixing between the synchrotron plasma and the  thermal plasma, especially near
the hotspot where the two plasmas are well  separated by the bow shock (Carilli
et al. 1994). It is believed that the radio  jets evacuate the cluster medium
along their paths with very little mixing of the two media. A similar effect is
visible in the galaxy 0902+343 (z=3.4) with its line-emitting gas 
anticorrelating with the radio source (Carilli 1995). Even theoretically it is
not clear  if the hotspot field can be diffused into the region around it on a
timescale fast  enough to produce a deep Faraday screen surrounding the hotspot
before it moves  ahead substantially (i.e. within $10^6$ years).

Accordingly, one must investigate alternative mechanisms for the strong magnetic 
fields at high redshifts indicated by the Faraday observations (e.g. Athreya 
1998).

Finally, a  magnetic field efficiently amplified during early galactic
evolution has to influence the various dynamical processes in galactic disks.
The coupling of Parker (buoyancy), Jeans (gravitational) and Field (thermal)
instabilities leads to the formation of giant molecular clouds (Blitz and Shu
1980). On the other hand the collapse of the cloud cores is expected to be
slowed down by the presence of magnetic pressure. The interplay of these two
processes provides a kind of self-regulation mechanism which may influence
the star formation rate and the initial mass function (IMF) (Rees 1994). The
star formation rate is the input parameter for our fast dynamo, thus the
interplay of these processes controls the efficiency of the cosmic ray driven
dynamo.

\section{Conclusions}

All observations indicate that young galaxies are characterized
by phases of strong star formation. The radio-far-infrared-correlation
observed in numerous galaxies with completely different star formation
rates proves that the formation of stars is accompanied by efficient
production of cosmic rays. In other words, a strong starburst is equivalent
to strong magnetic fields generated within a few $10^7\div 10^8$ years and an intense
cosmic ray flux. Observations of old stars in the halo of the Milky Way
supports this picture. They contain significant amounts of spallation products (B and Be)
which are due to a cosmic ray flux in the young Galaxy which was considerably
higher than today. On top of that, the cosmic ray propagation at these
early phases of galactic evolution can only be understood if the magnetic
field strengths were higher than its current values.

If we transfer these findings to young galaxies we expect that these
objects contain many more cosmic rays than fully evolved galaxies.
In a series of papers, we have shown that cosmic rays can drive an efficient
galactic dynamo based on outflows from the star formation regions. The advantage of the CR-dynamo is twofold: First, it is very fast
on time scales of a few $10^7\div 10^8$ years, whereas conventional dynamos amplify
the disk field on timescales of gigayears (Camenzind \& Lesch 1994).
Second, the CR-dynamo thrives on the outflows produced by star formation activity,
whereas conventional dynamos are susceptible to inevitable disruption by outflows.
The stronger the disk activity, the faster and more effective is the CR-dynamo.

Thus, very young galaxies are strongly magnetized. We predict that observations
even at very high redshifts ($z>4$) will find many galaxies with high radio fluxes.

\begin{acknowledgements} We deeply thank Heinz V\"olk for  insightful
discussions  and very helpful advices. We thank the referee L.O.C. Drury for
his comments which improved the paper.  This work was supported by the Polish
Committee for  Scientific Research (KBN) through the grant PB PB
404/P03/2001/20. The presented work is continuation of a research program
realized by MH under the financial support of {\em Alexander von Humboldt
Foundation}.  \end{acknowledgements}

\bibliographystyle{unsrt}

\end{document}